# Asymptotically Non-Singular Extended Non-Dyonic Solutions of 't Hooft-Polyakov Monopole Violates Equations of Motion

K. Rasem Qandalji
Amer Institute
*P.O.Box 1386, Sweileh, 11910, JORDAN*

**Abstract.** we show that based on the general solution, given by Corrigan, Olive, Fairlie and Nuyts, in the region outside the monopole's core; the equations of motion in the Higgs vacuum (i.e. outside the monopole's core) will not allow asymptotically non-singular extended non-trivial non-Dyonic (including, also, all static) solutions of the 't Hooft-Polyakov monopole. In other words, unless the monopole's magnetic charge is shielded (by some mechanism), the Dirac string is inevitable asymptotically, in the region outside the monopole's core, for all non-Dyonic solutions that are admissible by the equations of motion. That we show that the non-dyonic solutions (based on Corrigan *et al*) will include all "admissible" static solutions and their gauge transform might be interpreted as that all admissible dyonic solutions (based on Corrigan et al) are composite solutions.

## 1. Introduction.

't Hooft-Polyakov monopole [1,2], is based on the Georgi-Glashow SO(3) model, [3]. It introduced the first monopole model with a non-trivial non-singular interior, as well as an exterior with a "non-singular" extended solution that looks asymptotically like Dirac's monopole without the "singular" string; where the Dirac string could be gauged or smoothed away only for those solutions corresponding to elements of the first homotopy group of the exact symmetry group, SO(2), which are trivial in the complete symmetry group of the model, SO(3).

The spectrum of the allowed magnetic charges [4,5,6] for the 't Hooft-Polyakov monopole (corresponding to solutions with smooth non-singular field configuration of the monopole's interior and exterior) is only half those predicted by the Dirac quantization condition; that is caused by the fact that $\Pi_1(SO(3)) = \mathbb{Z}_2$, and only half of the homotopically distinct loops in (the exact symmetry group) SO(2) are trivial in (the complete symmetry group) SO(3). The monopoles, corresponding to those loops in SO(2) which are not trivial in SO(3), will necessarily have a Dirac string that can't be gauged away; and those don't belong to the 't Hooft-Polyakov monopole's spectrums since they will lead to a singular interior.

In this work we will show that the equations of motion, in the Higgs vacuum region, derived from the general solution (in that region) of Corrigan, Olive, Fairlie and Nuyts [7], will not allow, for the non-Dyonic solutions of the 't Hooft-Polyakov monopole, the (familiar) smoothing or gauging away of the Dirac string, i.e. in particular, we show that the smoothing out of the Dirac string is not allowed for non-Dyonic solutions corresponding to loops in

SO(2) which are trivial in SO(3); where, (unlike the solutions corresponding to loops Lie SO(2) which are not trivial in SO(3) which forced the string on us for topological reason that initiated at the singular core); in this case on the other hand, the Dirac string is inevitable asymptotically due to restrictions or constraints derived from the equations of motion in the Higgs vacuum that picks this particular gauge as the only one admissible by the equations of motion there.

From the constraint in Eq.(20) below, we see that the equations of motion will be violated for non-Dyonic solutions unless either the magnetic charge at the core of the monopole is "shielded" or otherwise the Dirac singular string is declared as inevitable asymptotically (i.e. in the Higgs vacuum).

## 2. Lagrangian Analysis and Derivation of the Constraints.

Gauge freedom of a singular physical system implies that the Euler-Lagrange equations of motion of the fields, of the theory at hand, are not all independent; i.e., we should be able to construct identities out of these equations of motions [8,9],. These identities will be satisfied by combinations of some rows (columns) of the Hessian matrix too; where the Hessian matrix, $\mathcal{M}$, is defined as the second derivative of the (local) Lagrangian density, with respect to the velocities of all the fields in the theory, ($\xi_a, \xi_b, etc...$): $\mathcal{M} \equiv \left\| \dfrac{\partial^2 \mathcal{L}}{\partial \dot{\xi}_a \partial \dot{\xi}_b} \right\|$. Further independent combinations of some rows (columns) of the Hessian matrix might also satisfy even more identities than those satisfied by the equations of motion; and that should result in putting "constraints" on the fields available in the theory and their velocities. These constraints should always be satisfied on any genuine path allowed by the equations of motion.

Next we will apply the above algorithm to the 't Hooft-Polyakov monopole and derive some constraints of interest to us.

The model at hand, the 't Hooft-Polyakov monopole, is based on the Georgi-Glashow model [3] that consists of SO(3) gauge fields interacting with Higgs isotriplet, $\boldsymbol{\phi} = (\phi_1, \phi_2, \phi_3)$. The Lagrangian of this model is, (signature of flat metric $= -2$ ),

$$\mathcal{L} = -\frac{1}{4} G_a^{\mu\nu} G_{a\mu\nu} + \frac{1}{2} \mathcal{D}^\mu \boldsymbol{\phi} \cdot \mathcal{D}_\mu \boldsymbol{\phi} - V(\boldsymbol{\phi}) \; , \qquad (1)$$

where, $V(\boldsymbol{\phi}) = \dfrac{1}{4} \lambda (\phi_1^2 + \phi_2^2 + \phi_3^2 - a^2)^2$; and where $W_a^\mu$ is the gauge potential. The gauge field strength, $G_a^{\mu\nu}$, is given by: $\mathbf{G}^{\mu\nu} = \partial^\mu \mathbf{W}^\nu - \partial^\nu \mathbf{W}^\mu + ie[\mathbf{W}^\mu, \mathbf{W}^\nu]$, evaluated in the SO(3) Lie algebra. In this model, the SO(3) gauge group, generated by the three $T_a$'s, is broken spontaneously, by the degenerate non-vanishing vacuum, down to SO(2) [$\simeq$ U(1)] gauge group that is generated by the charge, $\boldsymbol{\phi}\cdot\mathbf{T}/a$. This unbroken generator annihilates the Higgs field outside the monopole's core, and associated with it, is a massless (long-range) gauge potential that can be identified outside the monopole's interior with regular Maxwell's electro-magnetic field.

The size of the 't Hooft-Polyakov monopole's interior, (i.e. the monopole's core), is estimated using the Compton wavelength of the massive gauge bosons associated with the two broken generators and the mass of the surviving Higgs particle.

The model's energy finiteness implies [10] that, asymptotically, [i.e. in the "Higgs vacuum" region outside the monopole's core,] there are two conditions, below, which must be satisfied and hence defines the Higgs vacuum:

$$\mathcal{D}^\mu \boldsymbol{\phi} \equiv \partial^\mu \boldsymbol{\phi} - e \mathbf{W}^\mu \wedge \boldsymbol{\phi} = 0 ; \qquad (2)$$

(i.e. the Higgs field is covariantly constant in that region,) and

$$\phi_1^2 + \phi_2^2 + \phi_3^2 - a^2 = 0 \; (\Rightarrow V(\boldsymbol{\phi}) = 0), \qquad (3)$$

i.e. (in Higgs vacuum) we have: $\boldsymbol{\phi}\cdot\delta\boldsymbol{\phi} = 0$, or equivalently, $\phi_a \partial^\mu \phi_a = 0, \; (\mu = 0,...,3)$ \qquad (3.a)

The "general" form of $\mathbf{W}^\mu$ in Higgs vacuum was given by Corrigan, Olive, Fairlie and Nuyts [7]:

$$\mathbf{W}^\mu = \frac{1}{a^2 e} \boldsymbol{\phi} \wedge \partial^\mu \boldsymbol{\phi} + \frac{1}{a} \boldsymbol{\phi} A^\mu , \qquad (4)$$

where $A^\mu$ is arbitrary.

It follows that in the Higgs vacuum we have:

$$\mathbf{G}^{\mu\nu} = \frac{1}{a} \boldsymbol{\phi} F^{\mu\nu} ; \qquad (5)$$

where, 
$$F^{\mu\nu} = \frac{1}{a^3 e} \boldsymbol{\phi} \cdot (\partial^\mu \boldsymbol{\phi} \wedge \partial^\nu \boldsymbol{\phi}) + \partial^\mu A^\nu - \partial^\nu A^\mu . \qquad (6)$$

So in "Higgs vacuum"[use Eqs.(2, 3)]:

$$\mathcal{L} = -\frac{1}{4} G_a^{\mu\nu} G_{a\mu\nu} ;$$

and on account of (5, 6) we get,

$$\mathcal{L} = -\frac{1}{4} F^{\mu\nu} F_{\mu\nu}$$
$$= \frac{-\varepsilon_{ijk}\varepsilon_{rst}}{4a^6 e^2} \phi_i \phi_r \partial^\mu \phi_j \partial^\nu \phi_k \partial_\mu \phi_s \partial_\nu \phi_t - \frac{1}{2}\left(\partial^\mu A^\nu - \partial^\nu A^\mu\right)\partial_\mu A_\nu - \frac{\varepsilon_{ijk}}{a^3 e} \phi_i \partial^\mu \phi_j \partial^\nu \phi_k \partial_\mu A_\nu \qquad (7)$$

The Euler-Lagrange Equations of motion of our system in the Higgs vacuum, [see Eqs.(2-7)]:

$$0 = \frac{\delta S}{\delta \phi_m(x)} \equiv \frac{\partial \mathcal{L}}{\partial \phi_m(x)} - \partial^\sigma \frac{\partial \mathcal{L}}{\partial \partial^\sigma \phi_m(x)} = \frac{\varepsilon_{mjk}}{a^3 e}\left(\phi_j \partial^\mu \phi_k \partial^\nu F_{\mu\nu} - \frac{3}{2} F_{\mu\nu} \partial^\mu \phi_j \partial^\nu \phi_k\right),$$

$$0 = \frac{\delta S}{\delta A^\nu(x)} \equiv \frac{\partial \mathcal{L}}{\partial A^\nu(x)} - \partial^\mu \frac{\partial \mathcal{L}}{\partial \partial^\mu A^\nu(x)} = \partial^\mu F_{\mu\nu}.$$

(8)

Equations of motion, Eqs.(8), can be re-arranged [8,9] in order of time derivatives of the fields: i.e., the second time derivatives of the fields multiplying the elements of the Hessian matrix and the rest of the terms of lower time derivatives grouped in, call them, $\mathcal{K}_{\phi_m}$'s, and $\mathcal{K}_{A^\nu}$'s :

$$\left.\begin{array}{l} 0 = \dfrac{\delta S}{\delta \phi_m(x)} \equiv \mathcal{K}_{\phi_m}(\phi_l, \partial^\sigma \phi_n, A^\eta, \partial^\nu A^\mu) - \mathcal{M}_{\phi_m \phi_h}\ddot{\phi}_h - \mathcal{M}_{\phi_m A^\mu}\ddot{A}^\mu \\[1em] 0 = \dfrac{\delta S}{\delta A^\nu(x)} \equiv \mathcal{K}_{A^\nu}(\phi_l, \partial^\sigma \phi_n, A^\eta, \partial^\nu A^\mu) - \mathcal{M}_{A^\nu \phi_h}\ddot{\phi}_h - \mathcal{M}_{A^\nu A^\mu}\ddot{A}^\mu \end{array}\right\}$$

(9)

where,

$$\mathcal{K}_{\phi_m} = \frac{\varepsilon_{mjk}}{a^3 e}\left[-\frac{3}{2} F_{\mu\nu}\partial^\mu \phi_j \partial^\nu \phi_k + \phi_j \partial^\mu \phi_k \partial^i F_{\mu i} + \phi_j \partial^i \phi_k\left(\partial_i \dot{A}_0 - \frac{\varepsilon_{rst}}{a^3 e}\phi_r \dot{\phi}_s \partial_i \dot{\phi}_t\right)\right];$$

$$\mathcal{K}_{A^0} = \partial^i F_{i0};$$

$$\mathcal{K}_{A^j} = \partial^i F_{ij} - \partial_j \dot{A}_0 + \frac{\varepsilon_{rst}}{a^3 e}\phi_r \dot{\phi}_s \partial_j \dot{\phi}_t;$$

(10)

and, using Eq.(7), the elements of the (symmetric) Hessian matrix, $\mathcal{M}$, are:

$$\mathcal{M}_{\phi_l \phi_h} = \frac{\partial^2 \mathcal{L}}{\partial \dot{\phi}_l \partial \dot{\phi}_h} = -\frac{\varepsilon_{lmn}\varepsilon_{hrt}}{a^6 e^2}\phi_m \phi_r \partial^k \phi_n \partial_k \phi_t; \quad \mathcal{M}_{\phi_l A^j} = \frac{\partial^2 \mathcal{L}}{\partial \dot{\phi}_l \partial \dot{A}^j} = \frac{\varepsilon_{lmn}}{a^3 e}\phi_m \partial_j \phi_n;$$

$$\mathcal{M}_{\phi_l A^0} = \frac{\partial^2 \mathcal{L}}{\partial \dot{\phi}_l \partial \dot{A}^0} = 0; \quad \mathcal{M}_{A^\mu A^0} = \frac{\partial^2 \mathcal{L}}{\partial \dot{A}^\mu \partial \dot{A}^0} = 0; \quad \mathcal{M}_{A^i A^j} = \frac{\partial^2 \mathcal{L}}{\partial \dot{A}^i \partial \dot{A}^j} = -g_{ij}.$$

(11)

From the above Hessian matrix, Eqs.(11), and in addition to others, we get the following "independent" identities (of interest to us), where summation over repeated indices is understood:

$$\left.\begin{array}{l} \dfrac{\partial^2 \mathcal{L}}{\partial \dot{\phi}_l \partial \dot{\phi}_h} + \left(\dfrac{\varepsilon_{lmn}}{a^3 e}\phi_m \partial^k \phi_n\right)\dfrac{\partial^2 \mathcal{L}}{\partial \dot{A}^k \partial \dot{\phi}_h} \equiv 0, \\[1em] \dfrac{\partial^2 \mathcal{L}}{\partial \dot{\phi}_l \partial \dot{A}^j} + \left(\dfrac{\varepsilon_{lmn}}{a^3 e}\phi_m \partial^k \phi_n\right)\dfrac{\partial^2 \mathcal{L}}{\partial \dot{A}^k \partial \dot{A}^j} \equiv 0, \\[1em] \text{for } l = 1,2,3; \text{ and } \forall h\,(=1,2,3); \forall j\,(=1,2,3) \end{array}\right\} ;$$

(12)

Directly off Eqs.(12), we read the elements of the three independent zero-eigenvectors of the Hessian matrix, call them, $u^{\phi_m}{}_{(l)}$, $u^{A^\mu}{}_{(l)}$; ($l = 1,2,3$). [Symbolically; the identities are written as: $u^a{}_{(l)} \mathcal{M}_{ab} \equiv 0$]:

$$u^{\phi_m}{}_{(l)} = \delta_{ml}; \quad u^{A^k}{}_{(l)} = \frac{\varepsilon_{lmn}}{a^3 e} \phi_m \partial^k \phi_n; \quad u^{A^0}{}_{(l)} \text{ is arbitrary,} \quad (for\ l = 1,2,3). \quad (13)$$

If we multiply the equations of motion, Eq.(9), by the vectors $u_{(l)}$, Eqs.(13), we will, then, identically eliminate the second time derivative part in Eqs.(9); So, on genuine trajectories (i.e. where equations of motion are satisfied), Eqs.(9,12,13) lead to:

$$0 = u^a{}_{(l)} \mathcal{K}_a = \delta_{lm} \mathcal{K}_{\phi_m} + \left[\frac{\varepsilon_{lmn}}{a^3 e} \phi_m \partial^k \phi_n\right] \mathcal{K}_{A^k} + u^{A^0}{}_{(l)} \mathcal{K}_{A^0}, \quad (for,\ l = 1,2,3.) \quad (14)$$

Since $u^{A^0}{}_{(l=1,2,3)}$ is arbitrary, [see Eq.(13)], we (purposely) pick: $u^{A^0}{}_{(l=1,2,3)} = \frac{\varepsilon_{lmn}}{a^3 e} \phi_m \partial^0 \phi_n$, and using Eqs.(10), then, on genuine trajectories: Eq.(14) reduces to

$$0 = u^a{}_{(l)} \mathcal{K}_a = -\left(\frac{3}{2}\right) \frac{\varepsilon_{ljk}}{a^3 e} F_{\mu\nu} \partial^\mu \phi_j \partial^\nu \phi_k, \quad (for,\ l = 1,2,3.) \quad (15)$$

We form the following linear combinations of Eqs.(15), (or in other words; we form new vectors, $v^a_{(k)}$ and $w^a$, from the original $u^a_{(l)}$'s), that result in two independent identities, [Eqs.(16) below; these will be as shown to be identities on account of Eq.(3.a)]; as well as one constraint, [Eq.(17) below], that is independent of the two identities, [Eqs.(16)], and has to vanish (due to equations of motion) on genuine trajectories of the system. Eqs.(15) will, now, be (equivalently) replaced by the two identities, Eqs.(16), and the constraint, Eq.(17):

$$0 = v^a_{(m)} \mathcal{K}_a = \varepsilon_{mil} \phi_i u^a_{(l)} \mathcal{K}_a = -\varepsilon_{mil} \varepsilon_{ljk} \left(\frac{3}{2a^3 e}\right) F_{\mu\nu} \phi_i \partial^\mu \phi_j \partial^\nu \phi_k, \quad (for,\ m = 1,2) \quad (16)$$

where, $v^a_{(m)} \equiv \varepsilon_{mil} \phi_i u^a_{(l)}$.

We also have the constraint:

$$0 = w^a \mathcal{K}_a, \quad . \quad (17)$$

where, $w^a = \phi_l u^a_{(l)}$.

## 3. Dirac string is inevitable Asymptotically for non-Dyonic Solutions or Else

[First, recall that all our work here: the results and the derived constraints in sec.2. are (all) concerned with investigating the monopole's macroscopic field, or, the monopole's outer region (i.e. the Higgs vacuum). Also note that our (microscopic) monopole's core is, by

construction, smooth, non-trivial, and of finite size; as is required by the energy finiteness condition]

From Eq.(17), and using Eq.(15), we get the following constraint (in the Higgs vacuum) that have to be satisfied on any genuine trajectory admissible by the equations of motion in that region,

$$0 = \frac{\varepsilon_{ijk}}{a^3 e} F_{\mu\nu} \phi_i \partial^\mu \phi_j \partial^\nu \phi_k \equiv \frac{1}{a^3 e} F_{\mu\nu} \left[ \boldsymbol{\phi} \cdot (\partial^\mu \boldsymbol{\phi} \wedge \partial^\nu \boldsymbol{\phi}) \right], \tag{18}$$

or rewritten as,

$$\frac{2}{a^3 e} F_{0k} \left[ \boldsymbol{\phi} \cdot (\partial^0 \boldsymbol{\phi} \wedge \partial^k \boldsymbol{\phi}) \right] = -\frac{1}{a^3 e} F_{ij} \left[ \boldsymbol{\phi} \cdot (\partial^i \boldsymbol{\phi} \wedge \partial^j \boldsymbol{\phi}) \right], \tag{19}$$

From Eq.(19), and using Eq.(6) for $F^{\mu\nu}$ in the Higgs vacuum, we get:

$$\frac{2}{a^3 e} F_{0k} \left[ \boldsymbol{\phi} \cdot (\partial^0 \boldsymbol{\phi} \wedge \partial^k \boldsymbol{\phi}) \right] = -F_{ij} \left[ F^{ij} - \left( \partial^i A^j - \partial^j A^i \right) \right]. \tag{20}$$

We observe that, (in the Higgs vacuum), the left hand side of Eq.(20) will vanish, in a gauge-invariant way, only if the (gauge-invariant) electric field components, $F_{0k}$'s, vanish there; [i.e. (using Gauss theorem) for non-dyonic solutions.]

Note that the factor, $\left[ \boldsymbol{\phi} \cdot (\partial^0 \boldsymbol{\phi} \wedge \partial^k \boldsymbol{\phi}) \right]$, (appearing on the left hand side of Eq.(20)), is not gauge-invariant; and also note that we can always find some "gauge" in which this factor doesn't vanish. In fact, this factor does vanish in the temporal gauge (i.e., $\mathbf{W}^0 = 0$): where in that gauge we have, $\partial^0 \boldsymbol{\phi} = 0$, [as can easily be seen from Eqs.(3, 4); since the two terms on the right hand side of Eq.(4) are orthogonal to each other, and since we also have $|\boldsymbol{\phi}|^2 = a^2 \neq 0$ (in the Higgs vacuum)]; but, for any non-trivial Higgs configuration, we can always find some (time-dependent) gauge transformation that violates the temporal gauge, such that, in this new gauge, the factor, $\boldsymbol{\phi} \cdot (\partial^0 \boldsymbol{\phi} \wedge \partial^k \boldsymbol{\phi})$, does not vanish.

In the Higgs vacuum, and when the (gauge-invariant) $F_{0k}$'s vanish (on the left hand side of Eq.(20)), then the right hand side is required to vanish there and for "all gauges" too; and this implies that, we either have $F^{ij} = 0$ in the Higgs vacuum [i.e. the magnetic charge is shielded outside the monopole's core, by some mechanism]; or otherwise we must have the "gauge-dependent" statement that, (in that region), we have [compare with Eq.(6)]: $F^{ij} = \partial^i A^j - \partial^j A^i$, (i.e. only this particular "gauge" is allowed by the constraint of Eq.(20), which was forced on us by the equations of motion there in the Higgs vacuum). Then, in this latter case, the "singular" Dirac string will necessarily be reproduced outside the monopole's core (if we are going, at all, to have an asymptotically non-vanishing extended magnetic monopole field solution.), and this Dirac string, (by the equations of motion), is not allowed to be smoothed (or gauged) out in the Higgs vacuum.

By Gauss theorem; the $F_{0k}$'s vanish (in the Higgs vacuum) only for non-dyonic solutions. We also find (see below) that the $F_{0k}$'s vanish for "all" static solutions; (where recall again that our solutions here in the Higgs vacuum are based on the general solution given by Corrigan, *et al* see Eqs.(4,5,6)). The $F_{0k}$'s vanish also for any gauge transform of the static solutions since the $F_{0k}$'s are gauge-invariant.

Now we shall see that the non-dyonic solutions will include all the "admissible" static solutions and their gauge transform (since the $F_{0k}$'s are gauge invariant): That this is true can be seen (using the Gauss theorem and) since the $F_{0k}$'s vanish, [see Eq.(6)], for any "static" solution. To see that the $F_{0k}$'s vanish for static solutions (and their gauge transform) we need to use the fact that for any admissible (Corrigan *et al*) solution we must have $\partial^\mu A^0 = 0$, for all $\mu$; and this is true because the $A^0$ degree of freedom will be discarded, (so that the Poisson bracket may not be violated), on account that the $A^0$'s canonical momentum, $\Pi_0 \equiv \frac{\partial \mathcal{L}}{\partial \dot{A}^0}$, always vanishes (as can be seen from Eq.(7), of the Corrigan *et al* solution in the Higgs vacuum region).

That the non-dyonic solutions (based on Corrigan *et. al*) will include all admissible static solutions and their gauge transform will be interpreted (see below in the concluding section) as that the (Corrigan et al) dyonic admissible solutions are composite solutions and not elementary ones

Note that, for "static" solutions, we must also have $\partial^0 A^\mu = 0$, $\forall \mu$ (where this will also be substituted in Eq.(6) when computing the $F_{0k}$'s): To see that, recall first that for static solutions we must have, by definition, a "static" Higgs field configuration, (i.e., $\partial^0 \boldsymbol{\phi} = 0$), as well as, "static" gauge fields, (i.e. $\partial^0 \mathbf{W}^\mu = 0$). So, in particular, the component of $\partial^0 \mathbf{W}^\mu$ in the Higgs field (iso-space) direction should vanish for static solutions, and by using Eq.(4) we get:

$$0 = \boldsymbol{\phi} \cdot \partial^0 \mathbf{W}^\mu = \boldsymbol{\phi} \cdot \partial^0 \left[ \frac{1}{a^2 e} \boldsymbol{\phi} \wedge \partial^\mu \boldsymbol{\phi} + \frac{1}{a} \boldsymbol{\phi} A^\mu \right] = \frac{1}{a^2 e} \boldsymbol{\phi} \cdot \left( \partial^0 \boldsymbol{\phi} \wedge \partial^\mu \boldsymbol{\phi} \right) + a \left( \partial^0 A^\mu \right),$$

where the first term on the right hand side vanishes here since $\partial^0 \boldsymbol{\phi} = 0$ for static solutions, and hence (since the left hand side vanishes) the second term on the right hand side must also vanish (for static solutions), i.e. $\partial^0 A^\mu = 0$, $\forall \mu$.

## 4. Conclusion

Based on the most general solution given by Corrigan, Olive, Fairlie and Nuyts of the 't Hooft-Polyakov monopole in the Higgs vacuum region (i.e. outside the monopole's core): we showed above that one of the constraints derived from (the Lagrangian approach to) the equations of motion, [see Eq.(20)], will not allow (asymptotically) "smoothing" away of the singular Dirac string, for non-dyonic solutions (i.e. for which the $F_{0k}$'s vanish in Higgs vacuum). Here the equations of motion will not allow us to smooth the Dirac string in other SO(3) directions as is familiar to us and as always used to be done for monopole's with cores with topologically non-trivial but smooth Higgs field structure. The singular string survives asymptotically because the equations of motion forced this particular gauge with singular Dirac string as the only asymptotically non-vanishing allowed solution.

The only other alternative to save the constraint of Eq.(20) from being violated, is to have the magnetic charge (located at the core) shielded in the region outside the monopole's core, i.e. that the magnetic field vanishes in the monopole's outer region (i.e., in the Higgs vacuum). By the non-trivial construction of Higgs fields inside the monopole's core; we do know that we definitely have a magnetic charge at the core; and any claim of magnetic charge shielding needs further invistigation. So all we conclude here, at this point, is that the shielding of the monopole's magnetic charge for non-dyonic solutions in the Higgs vacuum is "consistent" with the constraints derived from the equations of motion and it is the only alternative to allowing only the solutions with singular Dirac strings asymptotically.

We also concluded above (based on equations of motion derived from the Corrigan, Olive, Fairlie and Nuyts solution in the Higgs vacuum), that non-dyonic solutions will include all "admissible" static solutions or any solution that can be gauge transformed into a static solution by some (gauge dependent) transformation.

What is interesting about finding that non-dyonic solutions will include all admissible "static" solutions (based on Corrigan *et al*) is that this might be interpreted as that the Corrigan *et al* solution predicts that all "dyonic" solutions based on theirs are composite particle solutions and not elementary: This can be seen to be true since (see [10]) for any "elementary" monopole solution we can always find a Lorentz frame and a certain (in general, time-dependent) gauge such that all the fields in our solution are static.

## Acknowledgment.
I thank the ILFAT Foundation (Ed'Oreen- Btouratij) for continuous support.